# Improving the quality of individual-level online information tracking: challenges of existing approaches and introduction of a new content- and long-tail sensitive academic solution

Silke Adam*, Mykola Makhortykh*, Michaela Maier**, Viktor Aigenseer***, Aleksandra Urman****, Teresa Gil Lopez*****, Clara Christner******, Ernesto de León******* and Roberto Ulloa********

\* Institute of Communication and Media Studies, University of Bern

\*\* Institute for Communication Psychology and Media Education, Rheinland-Pfälzische Technische Universität Kaiserslautern-Landau

\*\*\* AUNOVIS GmbH

\*\*\*\* Social Computing Group, University of Zurich

\*\*\*\*\* Department of Communication, University Carlos III of Madrid

\*\*\*\*\*\* Sustainability Office, Karlsruhe Institute of Technology

\*\*\*\*\*\*\*Amsterdam School of Communication Research, University of Amsterdam

\*\*\*\*\*\*\*\* University of Konstanz and GESIS – Leibniz Institute for the Social Sciences

## Author Note

**Improving the quality of individual-level online information tracking: challenges of existing approaches and introduction of a new content- and long-tail sensitive academic solution**

**Abstract:** This article evaluates the quality of data collection in individual-level desktop information tracking used in the social sciences and shows that the existing approaches face sampling issues, validity issues due to the lack of content-level data and their disregard of the variety of devices and long-tail consumption patterns as well as transparency and privacy issues. To overcome some of these problems, the article introduces a new academic tracking solution, WebTrack, an open source tracking tool maintained by a major European research institution. The design logic, the interfaces and the backend requirements for WebTrack, followed by a detailed examination of strengths and weaknesses of the tool, are discussed. Finally, using data from 1185 participants, the article empirically illustrates how an improvement in the data collection through WebTrack leads to new innovative shifts in the processing of tracking data. As WebTrack allows collecting the content people are exposed to on more than classical news platforms, we can strongly improve the detection of politics-related information consumption in tracking data with the application of automated content analysis compared to traditional approaches that rely on the list-based identification of news.

*Keywords:* online tracking, automated content analysis, WebTrack, content, long tail consumption.

**Introduction**

Individual-level online information tracking is a research method that seeks to capture individual media diets in online information environments. It allows researchers to observe who is exposed to what content on which websites and how individuals engage with the content to which they are exposed. In political communication research, individual-level tracking shows how people find political information online (de León et al., forthcoming; Möller et al., 2020a), what drives their information selection regarding ideology or demographics (Guess, 2021; Stier et al., 2020b), and which consequences such information yields on their attitudes and behaviors (Bach et al., 2021; de León et al., 2023; Guess et al., 2021).

In contrast to aggregate-level tracking, which captures information about media audience and itss overlap between media outlets (Dvir-Gvirsman et al., 2016), individual-level tracking focuses on personal media diets. It allows linking the volume and type of media usage to individual survey data (Stier et al., 2020a), which can reveal the drivers and outcomes of media diets. In addition, individual-level tracking can unravel personalized content exposure in an algorithmically curated media environment (Beam & Kosicki, 2014). It also allows including multiple platforms, such as social media or news websites—even those that are locked behind the paywall or registration (Loecherbach & Van Atteveldt, 2022).

Individual-level tracking is a reaction to the rise of a multi-channel online information environment, in which people can choose among a wide variety of channels spreading political information, entertainment, and misinformation and in which algorithms personalize what people see. It addresses the limitations of classical survey research, which fails to produce reliable measures of online information exposure, as surveys can only cover a fraction of the available online sources (González-Bailón & Xenos, 2022) and are subjected recall, social desirability and inattentive responding issues (Jürgens et al., 2019; Parry et al., 2021; Scharkow, 2016).

However, as research on (individual) online information tracking matures, it is clear that it suffers from severe data collection quality challenges (Jürgens et al., 2019; Parry et al., 2021, Bosch & Revilla, 2022). In identifying these quality challenges and developing a solution to some of them, we concentrate on desktop tracking, as it is more widely used in academic research and because tracking mobile devices, especially apps, is more challenging due to legal and ethical issues. To do so, we follow three research goals:

1) identify core aspects of webtracking data collection quality based on prior work and use these to evaluate the quality of existing desktop tracking data collections in academia;

2) introduce a new academic tracking solution that overcomes some of the major limitations of existing approaches in terms of data collection quality.

3) show in an empirical use case how the proposed solution for tracking data collection opens up new ways of improving the quality of tracking data processing: by moving beyond source-level analysis (e.g. identifying visits to news websites) and by adding the content-level analysis, we can improve the validity of studying different forms of information behavior (e.g. political information consumption) by systematically bringing in automated content analysis.

By improving the quality of data collected in web tracking research, namely by collecting what people actually see and do online on all platforms and therefore in the long tails of consumption patterns, we can also improve the quality of data processing which results in an increase in validity. Studying the content and the long tail of information consumption moves communication research beyond the analysis of (traditional) news and thus treats the online realm as more than an extension of the offline mass media. Beyond, adding the content to tracking data collections, we can bring in automated text analysis to distinguish whether content deals with political information or entertainment, refers to a specific ideology or contains specific arguments. Content analysis allows us to move beyond classifying news websites as quality or alternative and account for the diverse types of content present on these websites or on claiming that news readers consume the same content if they visit the same websites (Gentzkow & Shapiro, 2011).–We strengthen our argument by empirically illustrating that traditional source-based methods used to identify political information consumption in tracking data so far miss out dramatically compared to approaches that rely on automated content analysis.

**The quality of data collection in academic individual-level information tracking on desktop devices**

To evaluate the quality of data collection in academic desktop tracking research, we started from the classification of tracking error sources (Groves & Lyberg, 2010, Bosch & Revilla, 2022). We identified two main types of errors influencing tracking data collection: representation error (i.e. the failure to measure members of the population of interest shaping the sample quality) and measurement error (i.e.deviations between the concepts of interest and the measures collected shaping measurement quality). Following the error framework for webtracking by Bosch & Revilla (2022) and earlier reviews on the methodological approaches of tracking research (Christner et al., 2022; Loecherbach & Van Atteveldt, 2022), we discuss the *sample quality* and two aspects of measurement quality (the avoidance of *unobserved data* due to a sufficient tracking width and the *accurate specification* of concepts due to a sufficient tracking depth). The latter two quality criteria refer to the validity of data collection.

We argue, however, that the quality of data collection in webtracking is shaped – at least – by two further quality criteria. As tracking data potentially contains sensitive information leading to ethical challenges (Keusch et al., 2019; Makhortykh et al., 2022b), we regard *data privacy and minimization* as our fourth quality criteria (for a similar argument, see Bosch & Revilla, 2022). Finally, as hardly any standard exists concerning webtracking data collection transparency much research to date has not discussed it in detail (for a critique, Bosch & Revilla, 2022), we refer to the *transparency* as our final quality criteria.

These five criteria were used to evaluate data collection in individual academic desktop tracking research Hereby, we understand tracking solutions as technologies used for capturing "the traces that participants generate when interacting with their devices and online services" (Bosch & Revilla, 2022, p. 411). We look at data collections via commercial tracking solutions, as they are primarily used in the field (Christner et al., 2022) and via the few available academic open source tools. Regarding the latter, we focus on those tools that (1) "create data through recording behavior" (Loecherbach & Van Atteveldt, 2022), leaving out data donations that are dependent on the functionality of platforms from which the data stem and thus follow a different logic compared to other tracking approaches; that (2) capture real-world online usage instead of relying on tools simulating existing platforms, which causes validity issues as participants have to change their everyday routines; and that (3) provide open source solutions for tracking behavior on desktop devices. Considering the most recent tracking overviews (Christner et al., 2022; Loecherbach & Van Atteveldt, 2022), these criteria leave us with three open source academic trackers—Roxy (Menchen-Trevino & Karr, 2012), Eule (Haim & Nienierza, 2019), and Tracking exposed (Sanna et al., 2021)—in addition to commercial solutions offered by Wakoopa and RealityMine (that also provide these solutions to Netquest and YouGov).

*Sample quality*

Sampling issues have been widely discussed in survey research (e.g. Unangst et al., 2019), so we focus on sampling issues that are specific to tracking data collection. Tracking research can be based on probability samples or opt-in online panels of market researchers, but for our knowledge there are no tracking studies based on probability samples (see also Bosch & Revilla, 2022). This is due to commercial solutions focusing on opt-in samples and academic solutions' limited financial capabilities. The lack of probability samples also leads to a second sampling-related quality issue: if research primarily relies on the tracking samples of commercial companies already in place and more researchers use these samples, then potential biases in these opt-in samples are widely circulated. Although, one could just invest in a commercial solution and transfer it to another sample, due to financial constraints it is usually not a realistic option. In contrast, academic open source solutions can be more easily implemented in a larger variety of samples provided by market research companies or crowd-sourced. In addition to classic sampling issues, tracking studies face additional selection biases (Jürgens et al. 2019). These biases are related to specific segments of the population willing to be tracked due to it requiring more data sharing and more technical capabilities and raising more privacy concerns compared to traditional surveys. Tracking research struggles with limited willingness to participate and high drop-out rates (Gil-López et al., 2023; Wojcieszak et al., 2021, Revilla et al. 2017), thus prompting the importance of increasing tools' usability and decreasing their intrusiveness for improving recruitment processes. Regarding usability, the browser-based trackers (Eule, Tracking exposed) perform better than the proxy-based solutions, i.e., Roxy and proxy-based commercial solutions (Christner et al., 2022; Loecherbach & Van Atteveldt, 2022).

Furthermore, tracking approaches differ with regard to sample matching between survey and tracking data. The challenge is to collect tracking data only from those participants who take part in the survey and vice versa. This requires both study components being connected to an individualized login (Christner et al., 2022), as several members of the household may use the same device for different purposes. There is hardly any information on sample-matching precision in published studies with an exception of the Facebook tracker Eule, which allows for the assignment of login credentials to participants. Several other studies mention this problem and either call for login-protected tracking tools (Dvir-Gvirsman et al., 2016) or circumvent missing login systems by focusing on single-person households (Araujo et al., 2017). A few studies (e.g. Revilla et al., 2017; Konitzer et al., 2021; Bosch & Revilla, 2022) discuss the matching procedure for commercial solutions, noting that they require participants to mark which individual is using the device in multi-person households and that it is still difficult to deal with devices which can be potentially shared withnon-participants.

*Measurement quality*

At the core of the measurement quality is data validity, which indicate whether the concept measured equals the concept of interest (Bosch & Revilla, 2022). Two aspects of measurement quality are of particular relevance for tracking data collection: 1) the avoidance of unobserved data by relying on an appropriate tracking width; and 2) enabling data accuracy by relying on an appropriate tracking depth.

First, we consider the **tracking width**, which determines how many platforms are included in the tracking data collectionand how many devices are tracked. The required tracking width varies depending on the research interest: If, for example, a researcher is interested in Facebook only, then a single platform data collection is sufficient. Yet, if researchers seek to understand broad media diets, then a single platform and single device tracker will produce missing data. A middle ground is tracking approaches that collect data from a variety of platforms but rely on the principle of allow lists that ignores data coming from non-listed platforms (Christner et al., 2022). Such an approach helps implement data minimization for data collection (discussed later), but it is insufficient for capturing visits to less-known pages and long-tail consumption patterns. On the device level such middle-ground approaches would, for example, track multiple desktop devices (e.g. laptops and personal computers), but no mobiles. The broadest tracking approach relies on the deny-list principle, which captures all types of platforms and content except those that are defined in advance (e.g., banking websites due to privacy concerns). The deny-list approach enables the study of the long tail of online consumption habits that is important considering that people with specific characteristics engage with the niche outlets (Guess et al., 2018) and such niche exposure goes beyond incidental exposure (González-Bailón & Xenos, 2022). On the device-level, the broadest approach would allow for the inclusion of the full range of desktop and mobile tracking.

Open source academic tracking solutions usually focus on single social media platforms such as Facebook (Haim & Nienierza, 2019) or YouTube(Sanna et al., 2021). These trackers rely on browser-based scraping, which further limits the tracking width with to certain set of browsers. Roxy (Menchen-Trevino & Karr, 2012) was conceptualized as a broader tracker; however, it fails to capture the breadth of online content today due to its focus on http-content. Commercial solutions are, to our knowledge, platform (and browser) agnostic, as most of them rely on proxy technology and potentially allow for the inclusion of long-tail consumption patterns. Beyond, commercial solutions allow for tracking across different devices – mobile and desktop. Yet, these technological possibilities are hampered by people's lack of willingness to install a tracker on all devices (see Revilla et al. 2017, Barthel et al., 2020).

Second, validity issues also arise as regards the **tracking depth**, which is similar to what Christner et al. (2022) describes as the types of information that can be collected by a tracking device. We can distinguish data collections that provide information on the domain visited (e.g., Spiegel.de), the URL visited (a specific page on Spiegel.de) and what content people see and how they engage with it. For the latter, the tracker needs to capture not only tracking visits but also web content (i.e. text and images from the visited websites; Liu, 2008). Researchers have underlined the need to increase tracking depth by tracking the content (Dvir-Gvirsman et al., 2016; Möller, Hameleers et al., 2020) and moving beyond counting visits and classifying sources on the domain level. This is primarily necessary if the research question goes beyond measuring clicks and visits, but seeks to understand exposure to specific ideas and topics. Here, domain-level analysis might be misleading if, for instance, general news consumption is conflated with political news consumption (González-Bailón & Xenos, 2022) or if political exposure is limited to news outlets despite many other online sources that may contain political information (discussed later in this paper). In addition, domain-level analysis might lead to severe inconsistencies in the data, as a right-wing speech on a right-wing page would be classified as right-wing, whereas the same speech on a public TV webpage would be regarded as mainstream news (Dvir-Gvirsman et al., 2016). Therefore, bringing in the content is crucial because the consequences of media exposure are also present at the content level. It is the topics we read about, the positions made visible and the frames put forward that influence our attitudes and behaviors.

In general, there are two ways in which content can be integrated into tracking research. First, researchers can retrospectively identify the content to which participants are exposed using URLs. Second, the tracker can directly capture content during the data collection. The latter option has two advantages: it results in less missing data (e.g. a quarter of the pages could not be scraped retrospectively; Guess et al., 2020) and, most importantly, it is needed for studying the algorithmic curation of personalized online content (Christner et al., 2022), for instance, individualized web search outputs or personalized political ads.

Thus far, academic research primarily relies on commercial tracking solutions (Christner et al., 2022), where researchers buy domain- or URL-level data (as content-level data are quite expensive). To make sense of these data, most researchers apply source-level analysis, for example, by using source lists to identify all news pages (Bach et al., 2021; Scharkow et al., 2020), all alternative / populist pages (Bach et al., 2021), or by attributing ideological leanings to the domains based on alignment scores (Fletcher et al., 2021; Wojcieszak et al., 2021). A few studies have brought in the content by retrospectively scraping the URLs (Dvir-Gvirsman et al., 2016; Guess et al., 2020; Richter & Stier, 2022; Stier et al., 2021; Wojcieszak et al., 2021). Academic open source trackers increasingly seek to move beyond the URL level by capturing content which the user is engaged with. This holds for Eule (Haim & Nienierza, 2019) and for Roxy (Menchen-Trevino & Karr, 2012). However, the latter tool focuses on content from websites which use an http protocol instead of https, and such websites are increasingly rare.

In summary, validity issues also arise as regards tracking depth. Depending on the research questions posed, domain classifications might be inaccurate. High-quality data collection therefore requires the appropriate level of tracking depth because for answering many research questions, a higher depth is required.Commercial solutions are able to provide data on all levels, but in practice, content-level data are expensive and researchers often cannot afford it. This means that researchers interested in the content  retrospectively scrape it. This procedure does not allow studying the algorithmic curation of personalized content.

*Quality as regards privacy and data minimization in tracking data*

Tracking approaches must be evaluated against **privacy and data minimization standards**. This is an ethical but also partially legal obligation (Bodo et al., 2017), as tracking approaches contribute to a surveillance society (Zuboff, 2019). The principle of data minimization means that only those data that are necessary for a specific study are collected. Studies that, for example, deal with online news consumption are well equipped if they collect tracking data based on predefined media allow lists. In contrast, studies that seek to understand exposure to alternative information, which is often found in the long tail of consumption patterns, require broader tracking approaches based on deny lists. Thus, ideally, a tracker would allow flexible settings that could be adjusted to the research questions. Finally, tracking approaches need to avoid the collection of private information. This requires excluding sensitive web content (e.g., health information, banking) and implementing selectors for social media to collect only publicly available information. Moreover, tracking approaches need to avoid the collection of information from a person's network, i.e., from persons that have not given their informed consent but are linked to a tracked person. To do so, social media selectors are important, as are login credentials, to start the tracking process. Data privacy is also strengthened if the tracking device has a privacy function that allows users to surf without being tracked.

The few open source academic tracking solutions have implemented privacy features, all of which have login options. Eule and Tracking rely on an allow-list approach, with the former also using social media selectors. In contrast, Roxy is based on deny-list logic, to which users can contribute. In the case of commercial tracking solutions, the information about privacy and data minimization is not always made available by the studies relying on these solutions. In some cases, (Richter & Stier, 2022; Stier, Kirkizh et al., 2020).just note that the commercial solutions usedmeet legal standards, such as the European General Data Protection Regulation. In other cases, however, studies (e.g. Torcal et al., 2023) report that commercial solutions (e.g. Wakoopa trackers) capture all URLs visited by participants, thus raising concerns about data minimization (at least on the level of data collection); similarly, they acknowledge challenges of working with shared devices, where online activity of non-participants can also be tracked without them providing consent (Bosch & Revilla, 2022).

*Quality as regards transparency of tracking data*

**The transparency of the tracking process** for the researchers, is essential for evaluating the quality of thedata collection process and making the results (potentially) reproducible. While transparency is also important from the user perspective, we do not elaborate on it for this study due to the difficulties of obtaining information about this particular perspective from existing studies using tracking tools. There are several dimensions of tracking transparency that must be accounted for and that are closely related to the aforementioned criteria. First, transparency is necessary for the tracking width. Does the tracker work with allow or deny lists, and what sources are included in these lists? How does the tracker deal with social media data? How does it deal with long-tail consumption? Second, transparency is required for the technical details of the tracking process (Möller, van de Velde et al., 2020). How long must a page be viewed to count as a visit? How does the tracker deal with multiple tabs? How stable is data collection, and how much data loss is there? Third, transparency is necessary for data privacy and minimization. Which privacy features are implemented and how does the tracking approach seek to live up to the principle of data minimization?

Most tracking studies today rely on commercial solutions (e.g. Wakoopa tracking solutions) which have been criticized for their limited transparency (Jürgens et al., 2019). Often, the code of the tracker is not available and the details of the tracking process (e.g. what are the allow- or deny-lists) remain unclear. A few recent studies (e.g. Bosch & Revilla, 2022; Torcal et al., 2023) demonstrate the potential for more transparent reporting of the use of commercial tracking solutions by offering more information about their functionality (e.g. the exact types of data captured, the degree of tracking undercoverage, and the way privacy concerns are addressed). However, despite the gradual improvement in terms of transparency, commercial solutions still provide less capacities for researchers to ensure the transparency of data collection; sometimes, these limitations are attributed to privacy concerns as noted by Torcal et al. (2023) who noted that it was not possible to acquire raw data from the tracking service provider, so instead guidelines for how variables are to be calculated were sent to the service provider. Under the condition of the lack of access to raw data, researchers have to rely on commercial data providers for measuring possible tracking errors (Bosch & Revilla, 2022) and the degree of awareness about possible technical problems of the tracking process and the implementation of privacy measures can vary depending on how eager researchers are to request such information from the data provider. A higher level of transparency is one of the major advantages of all open source academic solutions.

*Summing it up*

The academic process of collecting webtracking data is faced by multiple severe quality challenges – the most outstanding are: 1) sampling issues as regards opt-in samples, recurring samples, sample bias due to participation issues and correct sample matching, 2) measurement issues as regards tracking width which potentially produces omitted data by design; 3) measurement issues as regards tracking depth, which avoids that tracking move beyond simple visit count data and study the content people are exposed to; 4) privacy and 5) transparency issues.

## WebTrack[1]—a new academic tracking solution

To tackle these challenges, we introduce a new academic tracking solution, WebTrack, which was developed within an SNF-DFG grant, empirically tested in several projects, and is now being made open source[2] and further maintained by a leading European research institution (GESIS, Cologne). WebTrack is a multi-platform tracker that captures content and the long tail of online information consumption, allows the application in a variety of settings and constant monitoring of the tracking process (i.e., data loss, consistency of the tracking sample, tracker settings), and is sensitive to privacy and data minimization principles. We explain why we have opted for a browser-based approach, discuss the design principles, provide details on the tracker settings from the perspective of the researcher and participant, elaborate on the requirements of the backend infrastructure, and finally evaluate our own approach based on the developed evaluation schema for online tracking. We also discuss its limitations, which refer to the lack of mobile tracking, its applicability to only two browsers, and the potential reactivity of tracking participants.

---

[1] WebTrack has been developed within the SNF-DFG funded project "Reciprocal relations between populist radical-right attitudes and political information behavior" (Adam & Maier; SNF, 100001CL_182630/1; DFG, MA 2244/9-1) as a joint effort of multiple team members: Viktor Aigenseer, Mykola Makhortykh, Silke Adam, Michaela Maier, Aleksandra Urman, Clara Christner and Teresa Gil-Lopez. In the further development and maintenance of the tracker Roberto Ulloa, Itay Caspari and Ernesto de León joined the team. The development of this tracker would not have been possible without additional funding by the Institute of Communication Psychology and the Research Center KoMePol at the University of Kaiserslautern-Landau and the Institute of Communication Science at the University of Bern. In addition, the maintenance of the tracker was also supported by an ERC-grant to Shira Dvir Gvirsman (ERC-2015-StG - 680009_SNSNEWS) and the DFG-project "Tracking the effects of negative political communication during election campaigns in on- and offline communication environments" (Maier, Stier & Otto; MA 2244-10/1 und STI 731/3-1).

[2] The source code for the backend (https://gitfront.io/r/user-6760823/Xkff33k5rYKD/webtrack-server/) and frontend (https://gitfront.io/r/user-6760823/FGeoQFp11QgS/webtrack-extension/) implementations of WebTrack is openly available on GitHub.

**Rationale for choosing a browser-based tracking approach**

Our decision to opt for a browser-based scraping approach is attributed to the limited availability of existing online tracking solutions (for an overview, Christner et al., 2022; Loecherbach & Van Atteveldt, 2022), as well as specific requirements of our project, namely the maximizing of the tracking breadth and width under the condition of high transparency and loyalty to privacy and data minimization principles.

Browser-based scraping approaches use browser extensions to capture HTML content that users view in their browsers. Thus, they allow tracking depth while also being flexible to be applied to different platforms on desktop devices (tracking width). Allowing a combination of depth and width—at least regarding desktop devices—makes them distinct from proxy approaches. Proxy approaches, which reroute traffic to proxy servers for information extraction, are strong regarding their applicability to different devices and platforms (width); however, they normally fail to go beyond domain- and URL-level information (depth). While some exceptions exist, the only publicly available one (i.e., Roxy; Menchen-Trevino & Karr, 2012) is capable of tracking content only from http domains. We refrained from developing a screen-recording approach that relies on software that captures what is happening on the screens of user devices, as this approach raises severe privacy concerns. Although strong regarding tracking depth and width, screen recording is highly intrusive (for ways to avoid this issue, see Krieter & Breiter, 2018) and thus even more complex for participant engagement than browser-based approaches. In addition, we also refrained from simulated environment approaches that mimic online environments, such as social media platforms (Masur et al., 2021) and a personalized news ecosystem (Loecherbach & Trilling, 2020), as they are subject to data validity issues due to participants' need to adapt to an artificially created environment and to limited tracking width as a substantial amount of resources is required for cross-platform simulations. Additionally, data donations were not a solution for our research interest, as this approach rarely provides information about the content level.

**WebTrack design logic**

The design logic of our approach is based on the screen-scraping principle, which shares several similarities to screen-recording approaches. However, it does not record the screen of the user device but instead extracts html appearing in the Internet browser on the device's screen. The focus on behavior in the browser (and not all applications that the participant might open on the screen, as in the case of screen-recording approaches) limits the tracking width of our approach but, at the same time, minimizes the amount of potentially irrelevant data that is captured and makes it easier to protect participants' privacy. Furthermore, the integration of not only the allow-list option (i.e., tracking only a selected list of websites) but also the deny-list option (i.e., tracking everything except a selected list of websites) expands the potential tracking width and allows better mechanisms for controlling it.

To implement this approach, we designed a browser extension called WebTrack. We focused on the extension of the Chrome browser, which currently remains the most commonly used Internet browser worldwide (Statcounter, 2022). Because of the high compatibility of extensions for Chrome and Firefox, we also adapted the extension for the Firefox browser; however, for Safari, which is the second most popular browser worldwide, easy adoption was not possible due to the different extension architectures. We also focused on the development of extensions for desktop versions of browsers due to the expectation that browser-based behavior is more common for desktop devices than for mobile ones.

The intended purpose of WebTrack design was to enable maximum tracking depth by capturing information not only about URLs opened by the participants in their browsers but also the dynamic and static content of the visited pages, including not only HTML but also potentially image and video URLs. Furthermore, the tool was designed to record the time participants spent on individual pages, as well as their interactions with social media content, including likes, reposts and comments. Because the design of WebTrack enabled the retrieval of large data volumes, including the content viewed by users in their browsers, it was essential to consider high ethical standards. To this aim, the design logic integrated the principles of data minimization, privacy, confidentiality and transparency, as we discuss below.

To implement the data minimization principle, we compiled a list of websites that were not intended to be tracked. The selection of the websites was based on the sensitivity of the information, which could potentially be tracked, as well as the requirements of the project (i.e., the focus on politics-related information behavior). Consequently, the list of websites that were not tracked included websites of insurance companies, medical services, pornography, banks, messengers /email-services and e-commerce. Furthermore, we integrated the possibility of updating or replacing the lists depending on the individual project requirements in the tool's functionality. In addition, we established a login procedure for each participant, requiring a unique tracking ID to be entered before tracking was activated to minimize the possibility of other users (e.g., members of the participant's family) being tracked without their consent or awareness.

To protect participants' privacy, we used the above-mentioned measures for data minimization, as well as the possibility of temporarily disable tracking by having a private mode setting for WebTrack, in which participants' activities in the browser were not tracked. Additionally, we implemented a system of filters for social media platforms that were tracked. The purpose of these filters was to remove personally identifiable information from the four social media platforms that were tracked (Facebook, Twitter, Instagram, YouTube). Specifically, the filters aimed to remove information about usernames and private content (e.g., non-public channels or posts, as well as personal messages); additionally, parts of platforms where the presence of such content was particularly high (e.g., personal settings) were on the deny list. Finally, for data storage, the pseudoanonymization principle was implemented, in which different sets of IDs for tracking and survey data were used to prevent the matching of information from these two datasets without a special matching table, which was only available to the project PIs.

In terms of confidentiality, the tool had to enable the encrypted exchange of data between the browser extension and the server where the data were stored, as well as the encryption of the storage itself to minimize the possibility of a data breach. In addition to these built-in mechanisms, strict data access protocols were implemented to protect the confidentiality of the participants. The downside was the limited capability of releasing data to the research community (at least in the non-aggregate form).

Finally, for the transparency principle, we enabled a selection of mechanisms to make the process more transparent for the researchers. Specifically, the integration of deny and allow lists in the interface enabled researchers to control the tracking width, whereas the differentiation between capturing data on the content, URL and domain levels enabled transparent control over the tracking depth. Furthermore, from the very beginning of the design process, the tool was intended to be released open source, which would also enable the research community and the general public to review its internal functionality.

**WebTrack interface and settings**

To elaborate on the functionality of WebTrack, it is important to discuss what its front-end implementation looks like. In the case of the participants, the interface is simple and consists of two parts: the pre-registration view and the post-registration view. The pre-registration view becomes available after the participant loads the plugin in the browser, after which the plugin is inactive until the participant enters the unique ID provided by the project team. After entering the ID, the interface shifts to the post-registration view, where the participant can either activate private mode or log out from the plugin. In private mode, the participant remains logged in but is not tracked anymore and receives notifications from the plugin as reminders to deactivate private mode. If the participant logs out of the plugin completely, WebTrack becomes inactive and shifts back to the pre-registration view.

**WebTrack backend requirements**

The frontend functionality of WebTrack relies on the backend infrastructure, which consists of several components. Two components, which are essential for WebTrack's functionality, are the NodeJS server on which the tool is running and the SQL database for storing metadata associated with individual HTML files captured by the tool. Both the Node.js server and the SQL database must be deployed on a machine that has sufficient computational resources. For our project with approximately 1200 participants, we used the Linux-based virtual machine with 16 CPU cores, 32 GB RAM, and a 10 TB hard drive (5 TB were used for data collection and 5 TB were used for data backup).

The architecture of the backend infrastructure required for WebTrack is further complicated by the sensitive nature of individual-level tracking data. Under these circumstances, it is important to adhere to high data privacy standards, which is usually achieved via storage in an encrypted environment, particularly if the storage is enabled through the third-party service provider, which is the case with many cloud-based data storage services outside of universities. One possible implementation is to send HTML files retrieved via WebTrack directly to the encrypted drive, which is attached to the server hosting the tool. However, it results in an increased workload for the server due to internal file movement and files' writing to the encrypted drive. Such a workload might potentially interfere with server performance and cause data loss. Given peaks of tracking traffic, during which participants browsed more intensively (usually in the evening hours) and the extra server workload due to encryption, it was necessary to integrate a distributed in-memory mechanism for queuing server jobs (Redis) that prevented the server from being overloaded.

**Evaluating WebTrack against tracking data collection quality criteria**

After describing the logic behind the design of WebTrack and the implementation of the tool, we critically reviewed it against the evaluation framework of tracking data collection quality outlined earlier.

The major advantage of WebTrack is that it provides a high *tracking depth*. It captures information about user visits simultaneously on the URL and the content level. Furthermore, depending on the sensitivity of the content, it can be adapted to capture data from certain websites on the domain, URL or content level. Such flexibility enables possibilities for data collection that cannot be matched by openly available academic tools and only available via commercial tracking solutions. However, compared with commercial solutions, WebTrack provides the same control over tracking depth for the researcher without the necessity to rely on a private company and, potentially, for a lesser amount of money, in particular for content-level data. It also gives researchers access to raw data (which is sometimes not accessible via commercial companies due to privacy concerns; Torcal et al., 2023) and allows more possibilities for tracking different types of tracking errors (Bosch & Revilla, 2022).

WebTrack offers a potentially unlimited *tracking width as regards different platforms accessed via desktop browsers and tracked in the study*. However, one major limitation here relates to ethical considerations; that is, while capturing all platforms visited by the user is possible (by removing deny-listed websites), doing so might lead to severe issues related to data minimization and potential privacy breaches (see more below). Beyond, WebTrack offers only limited tracking width as regards devices. So far it only captures browser-based online behavior on desktop devices for two browsers only. This is a major limitation as research shows that mobile devices turn into major access points (Reuters Digital News Report, 2022) with all social groups spending the majority of their time on mobile devices (Festic et al., 2021). Although the tool can potentially be adapted for web browsers used on mobile devices, it would not be able to capture behavior happening in mobile apps. In this sense, WebTrack's major limitation is its limited tracking depth as regards mobile devices.

With regard to *tracking samples*, there is little information on how different tracking tools perform in terms of dropout rates for participants whose behavior is tracked. For WebTrack, the dropout rate turned out to be moderately high when used in our own project, which can be attributed to the usability aspects of the tool. From the sample of 3249 users who agreed to be tracked, 1185 users (i.e., ~37%) registered at least one visit with WebTrack. To install WebTrack, participants must acquire the plugin via the Google store (for Chrome) or add it manually to the browser (for Firefox), then open the plugin and log in. While it is still substantially easier than, for instance, setting up screen-recording or proxy-based tools, these actions require certain effort from the participants and seem to increase the underrepresentation of older citizens (in particular, females) and less educated citizens. In addition, Individualized login procedures allow for precise matching of tracking and survey participants. Finally, as WebTrack is an open source tool, it can be used for a large variety of samples.

Because of the high tracking width as regards platforms and depth, WebTrack poses a number of challenges for *data minimization and participant privacy*. To counter these challenges, the tool's design includes several mechanisms (both on the frontend and backend levels) to enable data minimization and safeguard participant privacy. The key to these mechanisms is the private mode, which enables participants to stop their behavior from being tracked, for instance, when they are involved in more sensitive forms of information behavior. In terms of data minimization, WebTrack easily allows deny and allow lists to be added or modified to filter out domains that are not relevant to the project. Furthermore, WebTrack contains custom-made filters for major social media platforms to avoid capturing private data. However, frequent changes in the architecture of social media platforms make it necessary to adapt the filters constantly to prevent them from losing their functionality.

In terms of *transparency*, WebTrack enables multiple possibilities for evaluating the quality of tracking data. It allows researchers to specify and control the tracking width (e.g., via allow and deny lists) and depth and to review the settings through the interface. It also makes it clear how potentially sensitive data (e.g., private social media posts) are handled through the combination of openly accessible code and tool documentation describing the customized filters. Another major advantage of WebTrack is its release as an open source tool on Github that further amplifies the transparency of tracking procedures, at least on the side of the academic community.

**Beyond: The performance and future of WebTrack**

Our experience with WebTrack's performance was positive overall. For the relatively large project of tracking 1185 participants for 2.5 months in spring 2020, the tool did not experience any major interruptions. At some point, the web server used for hosting the tool started experiencing substantial overload, partly attributed to the high volume of content coming from a few users for a short period of time at peak hours. However, the implementation of the Redis-based in-memory queue mechanism addressed this performance-related issue.

As for the future of the tool, it will be maintained by GESIS, a major European institution, as a publicly open infrastructure, as well as released as an open source research solution. The existence of such a publicly open infrastructure has been called for by researchers for a long time (González-Bailón & Xenos, 2022). To highlight the possibilities for political communication research provided by the tool, we discuss in more detail one of its possible use cases, i.e., the combination of individual-level tracking and automated content analysis, in the next section.

**The future of academic tracking research: Integrating automated content analysis into tracking (and survey) research**

Following the introduction of WebTrack, we present how two features of the tool - i.e. the collection of content-level data and of visits to domains not limited to news - can break up current limitations of political communication research. Improvements in data collection can lead to improvements in data processing and increase the validity of constructs measured. To do so, tracking data must not only be combined with surveys, but also with elaborate forms of automated text analysis. These innovative approaches are not necessary for answering all research questions, but are important for studies looking beyond sources, scrutinising topics, actors or frames and assuming that political communication is not limited to news media. This focus on content and long-tail consumption patterns can lead to a paradigm shift in exposure research away from the classification of sources and reliance on visit volume toward a finer-grained understanding of what people see and do online in a multi-channel information environment.

This focus on content is rare, as most tracking studies today still focus on sources and the number of visits and thus replicate with tracking data what we have studied in traditional survey-based studies on media usage. The few researchers who have considered content have focused on further classifying online news, e.g. the share of political news in the news sections (Wojcieszak et al., 2021), the consumption of different news types (Stier et al., 2021, Peterson & Damm 2019), the ideological leaning (Peterson et al. 2019) or specific issues or personas in the news sections (de León et al., 2023; Richter & Stier, 2022). Few studies have moved beyond news content by either hand-coding all top domains (Dvir-Gvirsman et al., 2016) or by using keyword searches, e.g., to identify vaccine-related information (Guess et al., 2020).

However, automated content classification of tracking data is challenging—primarily if we move beyond the news. Tracking data contain a high volume of noise, as it comes in different formats (e.g., videos, texts), different platforms (e.g., e-commerce platforms, web search engines, social media), different lengths (e.g., Twitter tweets versus journalistic reports), and different languages. Furthermore, there are few automated text classifiers readily available beyond English, and due to privacy reasons, tracking data cannot be coded by crowdsourced workers. However, despite these challenges, in the following section, we seek to illustrate how tracking research might profit from automated approaches for analyzing text content.

Therefore, we compared the following approaches to identify political information in tracking data: the classical list-based approaches that identify news sites (e.g. Bach et al. 2021, Fletcher et al. 2021, Möller et al. 2020, Scharkow et al. 2020, Stier et al. 2021), the more elaborate approaches that combine list-based approaches with the classification of political information on the identified news sites (e.g. Wojcieszak et al. 2021, Guess 2021, Flaxmann et al. 2016) and a new approach that classifies political information not only on news sites but on all potential information sources in the tracking data, such as social media, organizations' websites, and blogs.

Tracking data were collected from a sample of Internet users in the age range of 18–75 years from Germany and German-speaking Switzerland. Recruitment was conducted via a Demoscrope market research company using online access panels and was based on representative quotas for the respective countries regarding gender, age and education (for a detailed analysis of sample composition, see Gil-López et al., 2023). After giving informed consent, the participants installed WebTrack. Overall, 583 people in Germany and 602 in Switzerland participated in tracking from March 17 to May 26, 2020. During this tracking period, we collected 2,084,400 web visits in total, which excluded visits to websites dealing with insurance, medical services, pornography, banks, messengers /email-services, online surveys and e-commerce and visits to pages containing content not primarily in German (for detecting the language of the page, we used the langdetect Python package; Danilak, 2023). The above-mentioned visits to privacy-sensitive pages were filtered out using the deny-list approach.

The first approach to identify political information consumption used *lists of news websites*. This is the most widely used approach in existing research. For our tracking project, we assembled a list of German-speaking news websites that contain journalistic media (see Appendix A). The second approach also relied on the identification of news websites based on lists but considered that news contains not only politically relevant information, but also non-political information (e.g. on sports, entertainment and celebrities). To account for this complexity, the second approach combines a list-based procedure with the *training of a political classifier to distinguish political from non-political news* based on previous work (Makhortykh et al., 2022a; Guess, 2021; Stier et al., 2021). We defined political information as information referring to political processes and procedures (politics), structures and institutional aspects (polity), and / or the content and actors of political disputes (policy). To detect such information, we relied on a dictionary approach, which showed good performance in the detection of political information in journalistic texts (F1 = 0.76; Makhortykh et al., 2022a) based on a comparison with manually coded golden standards (for more information, see Appendix B). This dictionary was made up of terms from the German codebook of the Comparative Agendas Project (CAP; Bevan, 2019) and was enhanced with a list of terms on topics that are underrepresented in the CAP codebook (i.e., elections and ecology) together with a list of political actors' names in Germany / Switzerland, as well as in the EU / G20. The final dictionary can be found here (https://osf.io/6dxbu?view_only=0c58144e1769492cb32dd2d650062534). The third approach applied the *classifier beyond the identified news websites to the overall tracking data*. To judge the performance, we created a manually coded gold standard capturing content from a broad range of platforms present in the tracking data. The dictionary performed well, with macro average F1 scores of 0.84 and 0.82 archived by the dictionary used in production (Makhotykh et al., 2023; see Appendix B for more information).

Our results regarding the list-based classification of news domains are in line with what has been found previously. News does not dominate online behavior. Of all our tracking data, 8.39% comprised visits to news domains based on the first approach (i.e. the matching of domains visited by participants with the list of news domains). Wojcieszak et al. (2021) found only 1.6%. Using the second approach (i.e. applying the dictionary to distinguish political from non-political news to the content coming from news web pages visited by the participants) this share further goes down to 2.53%, which replicates what Wojcieszak et al. (2021) found; only a fraction of news consumed actually refers to political information.

This indicates that using news websites as a single indicator for political information exposure is an imprecise or even misleading proxy. This also clearly applies to the ideological slant, which is often attributed to a news source instead of the specific content items individuals are exposed to. Finally, if we look at the full universe of information exposure—namely visits to all tracked websites (and not only news ones)—using the third approach (i.e. the application of the classifier to the content of all pages tracked in the course of the project), then the share of politics in our tracking data increases to 11.98%. This clearly shows that continuing to focus only on political information consumption on news websites, instead of expanding to the diverse sources of information that are used online, hampers academic research. It becomes clear that political information is more than news and that we should start analyzing long-tail consumption patterns.

## Discussion

WebTrack can not solve all potential pitfalls of webtracking data collection. Its major limitation is that it only captures desktop usage based on two browsers and it leads to a limited tracking width. Consequently, it might serve as an addition to alternative tracking solutions or other means of getting information about in-app usage (e.g. experience sampling ). The benefits of WebTrack are that it directly captures content (and thus allows for more elaborate analyses), can be used to study online content beyond news and include the long-tail consumption patterns, facilitates work with probability samples, is open access, and more transparent (and cheaper) than commercial solutions. These points are important because via realtime screen-scraping WebTrack can reduce the resources necessary for retrospectively scraping the content while allowing to study algorithmic content curation. As regards tracking width, WebTrack offers functionality similar to the commercial solutions making it easier for academic projects to conduct content-level analyses and study long-tail consumption and enabling new possibilities to move towards probability sampling in computational research that is needed to increase data reliability and representativeness.

In addition, WebTrack offers full transparency; its code is open source, and the tracking width and depth can be determined for each project. It gives control over the technical challenges in the data collection process and allows for a wide variety of privacy / data minimization options. Finally, WebTrack is constantly maintained by GESIS and thus serves as a research infrastructure for the social sciences (for the necessity of such collaboration in developing and maintaining open source tools, see Loecherbach & Van Atteveldt, 2022).

However, there are still tracking data collection quality problems which are yet to be addressed. As already stated before, WebTrack is limited to browser tracking despitethe rising importance of mobile app-based information consumption (e.g. Reuters Digital News Report, 2022). Due to privacy concerns, WebTrack does not allow capturing private communication via social media or messaging apps. Additionally, WebTrack captures data from only two browsers, Firefox and Chrome, and thus excludes participants who use other browsers (e.g., Safari). Another limitation is that WebTrack takes only one snapshot of each visit, whereas modern websites support dynamic interactions with users. Furthermore, despite its relative ease of use, WebTrack still has usability issues: at all stages of the recruitment process (from general willingness to share tracking data and give informed consent to installing the tracker), people drop out (for a detailed analysis, see Gil-López et al., 2023). As others have already pointed out (Wojcieszak et al., 2021), tracking studies face the challenge of potentially biased samples, as those who are willing to share their data might not necessarily represent the overall population. Finally, there is the intrusiveness of tracking, which potentially changes the participants' typical online behavior. The degree to which this actually takes place is hard to judge, as it would require comparing tracked behavior to untracked behavior, which cannot be observed.

In our eyes, the collection of tracking data requires that we take into account content, long-tail consumption, and mobile and desktop devices. WebTrack succeeds in two of these requirements while missing out on the mobile dimension. However, it already allows us to show the potential of such fine-grained data in terms of linking the content to which individuals are exposed online to survey data by means of automated content analysis. This combination of tracking, survey and automated text analysis is where we see the true potential of tracking research. We are more doubtful whether tracking research can serve as gold standard against which survey data are evaluated (as many of the comparisons are hampered not only by imprecise survey but also by imprecise tracking data, see for a discussion of tracking data errors Revilla et al. 2017, Bosch & Revilla, 2022, Barthel et al., 2020; González-Bailón & Xenos, 2022; Parry et al., 2021). In our eyes, tracking data – if they do not meet all requirements as regards width and depth, should be treated as another proxy for online media exposure with its own flaws and limitations. The innovative aspect of tracking research is that we can determine what people actually see and do online and look at the content and interactive behavior on the full breadth of platforms and the long tails of consumption patterns. This allows us to move beyond studying "the overall duration of volume of usage" (Parry et al., 2021) and to consider content outside news domains (Wojcieszak et al., 2021), particularly in the long tails of media consumption.

IMPROVING THE QUALITY OF INDIVIDUAL-LEVEL ONLINE INFORMATION TRACKING					35Peterson, E. & Damm, E. (2019). A Window to the World: Americans' Exposure to Polistical News From Foreign Media Outlets. https://doi.org/10.31235/osf.io/er48b

Peterson, E.; Goel, S. & Iyengar, S. (2019). Echo Chambers and Partisan Polarization: Evidence from the 2016 Presidential Campaign. Stanford PACS.

Reuters Digital News Report (2022). https://reutersinstitute.politics.ox.ac.uk/digital-news-report/2022

Revilla, M., Ochoa, C., & Loewe, G. (2017). Using passive data from a meter to complement survey data in order to study online behavior. Social Science Computer Review, 35(4), 521-536. https://doi.org/10.1177/0894439316638457

Richter, S., & Stier, S. (2022). Learning about the unknown Spitzenkandidaten: The role of media exposure during the 2019 European Parliament elections. *European Union Politics*, *23*(2), 309–329. https://doi.org/10.1177/14651165211051171

Sanna, L., Romano, S., Corona, G., & Agosti, C. (2021). YTTREX: Crowdsourced analysis of YouTube's recommender system during COVID-19 pandemic. *Information Management and Big Data*, *1410*, 107–121.

Scharkow, M. (2016). The accuracy of self-reported Internet use - A validation study using client log data. *Communication Methods and Measures,* *10*(1), 13–27. https://doi.org/10.1080/19312458.2015.1118446

Scharkow, M., Mangold, F., Stier, S., & Breuer, J. (2020). How social network sites and other online intermediaries increase exposure to news. *Proceedings of the National Academy of Sciences of the United States of America*, *117*(6), 2761–2763. https://doi.org/10.1073/pnas.1918279117

Statcounter. (2022, December). *Browser market share worldwide*. Statcounter. https://gs.statcounter.com/browser-market-share

IMPROVING THE QUALITY OF INDIVIDUAL-LEVEL ONLINE INFORMATION TRACKING										36

**Appendix**

**Appendix A: List of German-speaking journalistic outlets used for political news detection use case**

aachener-nachrichten.de
aachener-zeitung.de
schwaebische.de
wiesbadener-kurier.de
bo.de
bnn.de
kreiszeitung.de
weser-kurier.de
wn.de
augsburger-allgemeine.de
aichacher-zeitung.de
swp.de
alfelder-zeitung.de
waz-online.de
allgaeuer-anzeigeblatt.de
idowa.de
azonline.de
allgemeine-zeitung.de
come-on.de
giessener-allgemeine.de
pnp.de
az-online.de
nordbayern.de
mittelbayerische.de
onetz.de
nordkurier.de
derpatriot.de
harlinger.de
svz.de
aerztezeitung.de
bz-berlin.de
bkz.de
nw.de
westfalen-blatt.de
rnz.de
fnp.de
badische-zeitung.de
badisches-tagblatt.de
shz.de
infranken.de
bayerische-staatszeitung.de
bayernkurier.de
berchtesgadener-anzeiger.de
rundschau-online.de

rp-online.de
rhoenundsaalepost.de
ostsee-zeitung.de
morgenweb.de
abendblatt-berlin.de
berliner-kurier.de
morgenpost.de
berliner-zeitung.de
noz.de
bild.de
szbz.de
bbv-net.de
boehme-zeitung.de
borkenerzeitung.de
boersen-zeitung.de
mainpost.de
main-echo.de
maz-online.de
braunschweiger-zeitung.de
brv-zeitung.de
boyens-medien.de
allgaeuer-zeitung.de
buerstaedter-zeitung.de
butzbacher-zeitung.de
staatsanzeiger.de
cellesche-zeitung.de
ovb-online.de
evangelisch.de
zeit.de
cnv-medien.de
merkur.de
echo-online.de
das-parlament.de
24vest.de
dewezet.de
n-land.de
freitag.de
prignitzer.de
tagesspiegel.de
teckbote.de
westallgaeuer-zeitung.de
deutsche-handwerks-zeitung.de
dvz.de
die-glocke.de
dieharke.de
nq-online.de
rheinpfalz.de
die-tagespost.de

welt.de
da-imnetz.de
mittelhessen.de
lvz.de
saechsische.de
donaukurier.de
dorstenerzeitung.de
dnn.de
dzonline.de
gea.de
kn-online.de
goettinger-tageblatt.de
einbecker-morgenpost.de
ejz.de
emderzeitung.de
ev-online.de
esslinger-zeitung.de
express.de
fehmarn24.de
fla.de
verlag-dreisbach.de
hna.de
wlz-online.de
frankenpost.de
faz.net
fr.de
flz.de
fnweb.de
freiepresse.de
fuldaerzeitung.de
gandersheimer-kreisblatt.de
gaeubote.de
gnz.de
kreis-anzeiger.de
general-anzeiger-bonn.de
ga-online.de
giessener-anzeiger.de
gifhorner-rundschau.de
gmuender-tagespost.de
goslarsche.de
gn-online.de
muensterschezeitung.de
haller-kreisblatt.de
halternerzeitung.de
abendblatt.de
mopo.de
hanauer.de
handelsblatt.com

haz.de
harzkurier.de
stimme.de
hellwegeranzeiger.de
helmstedter-nachrichten.de
hersfelder-zeitung.de
hildesheimer-allgemeine.de
hurriyet.com.tr
ivz-aktuell.de
ikz-online.de
juedische-allgemeine.de
kevelaerer-blatt.de
ksta.de
kornwestheimer-zeitung.de
krzbb.de
lahrer-zeitung.de
lampertheimer-zeitung.de
szlz.de
landeszeitung.de
ln-online.de
lr-online.de
lauterbacher-anzeiger.de
leinetal24.de
leonberger-kreiszeitung.de
lz.de
lkz.de
main-spitze.de
marbacher-zeitung.de
verlagshaus-jaumann.de
moz.de
medical-tribune.de
insuedthueringen.de
derwesten.de
milligazete.com.tr
mt.de
mz-web.de
tag24.de
muehlacker-tagblatt.de
muensterlandzeitung.de
om-online.de
mv-online.de
murrhardter-zeitung.de
maerkte-weltweit.de
naumburger-tageblatt.de
neckar-chronik.de
ndz.de
neuepresse.de
np-coburg.de

nrz.de
nordbayerischer-kurier.de
nnn.de
nord24.de
nwzonline.de
ntz.de
op-marburg.de
oberhessische-zeitung.de
obermain.de
wnoz.de
rhein-zeitung.de
op-online.de
oz-online.de
otz.de
paz-online.de
peiner-nachrichten.de
pz-news.de
pirmasenser-zeitung.de
remszeitung.de
rga.de
tagblatt.de
rheiderland.de
rheingau-echo.de
ruhrnachrichten.de
saarbruecker-zeitung.de
salzgitter-zeitung.de
sn-online.de
schifferstadter-tagblatt.de
zvw.de
schwaebische-post.de
schwarzwaelder-bote.de
schwarzwaelder-post.de
beobachter-online.de
serbske-nowiny.de
siegener-zeitung.de
soester-anzeiger.de
solinger-tageblatt.de
tageblatt.de
stuttgarter-nachrichten.de
stuttgarter-zeitung.de
sueddeutsche.de
suedkurier.de
tah.de
taz.de
thueringer-allgemeine.de
tlz.de
torgauerzeitung.com
traunsteiner-tagblatt.de

volksfreund.de
tz.de
uckermarkkurier.de
usinger-anzeiger.de
vkz.de
vdi-nachrichten.com
vesti-online.com
vogtland-anzeiger.de
volksstimme.de
wz-net.de
werra-rundschau.de
wz.de
wp.de
wr.de
wa.de
wetterauer-zeitung.de
lokal26.de
wolfenbuetteler-zeitung.de
wolfsburger-nachrichten.de
wormser-zeitung.de
yeniozgurpolitika.net
zak.de
spiegel.de
focus.de
stern.de
wiwo.de
manager-magazin.de
abendzeitung-muenchen.de
life-und-style.info
myself.de
ok-magazin.de
superillu.de
instyle.de
wunderweib.de
freizeitrevue.de
glamour.de
faces.ch
3sat.de
arte.tv
bb-mv-lokaltv.de
br.de
brf.de
daserste.de
dw.com
hr-fernsehen.de
kika.de
mdr.de
ndr.de

phoenix.de
radiobremen.de
rbb-online.de
sr.de
swrfernsehen.de
ard.de
wdr.de
zdf.de
deutschlandfunk.de
deutschlandfunkkultur.de
deutschlandradio.de
deutschlandfunknova.de
hr1.de
hr2.de
hr3.de
hr4.de
you-fm.de
hr-inforadio.de
n-joy.de
antennebrandenburg.de
fritz.de
inforadio.de
radioeins.de
unserding.de
dasding.de
swr3.de
wdrmaus.de
1-2-3.tv
anixehd.tv
astrotv.de
bibeltv.de
bwfamily.tv
channel21.de
comedycentral.tv
deluxemusic.tv
deraktionaertv.de
deutsches-musik-fernsehen.de
disney.de
dmax.de
drf-tv.de
eotv.de
euronews.com
eurosport.de
ewtn.de
www-health.tv
hgtv.com
hopechannel.de
hse24.de

juwelo.de
k-tv.org
kabeleins.de
kabeleinsdoku.de
mediashop.tv
mtv.de
n-tv.de
nick.de
nitro-tv.de
pearl.de
prosieben.de
prosiebenmaxx.de
qs24.tv
qvc.de
rictv.de
rtl2.de
rtlplus.de
rtl.de
sat1.de
sixx.de
sonnenklar.tv
sport1.tv
startv.com.tr
superrtl.de
tele5.de
tlc.de
toggo.de
vox.de
weltderwunder.de
xite.tv
sky.de
13thstreet.de
animalplanet.de
auto-motor-und-sport.de
aenetworks.de
axn.com
bongusto.de
cartoonnetwork.de
stingray.com
discovery.com
foxchannel.de
geo-television.de
goldstar-tv.de
gutelaunetv.de
heimatkanal.de
history.com
jukebox-tv.de
junior-programme.de

kabeleinsclassics.de
kinowelt.tv
lust-pur.tv
marcopolo.de
motorsport.tv
motorvision.tv
nationalgeographic.de
natgeotv.com
nauticalchannel.com
nickjr.de
planetradio.de
prosiebenfun.de
rck-tv.de
rtl-crime.de
rtl-living.de
rtl-passion.de
sat1emotions.de
silverline24.de
sonychannel.de
syfy.de
tnt-tv.de
universaltv.de
absolutradio.de
energy.de
erf.de
klassikradio.de
lulu.fm
schlagerradiob2.de
radiobob.de
horeb.org
schlagerparadies.de
radioteddy.de
rockantenne.de
hitradio-rtl.de
schwarzwaldradio.com
sunshine-live.de
antenne1.de
bigfm.de
egofm.de
radio7.de
regenbogen.de
baden.fm
dieneue1077.de
die-neue-welle.de
donau3fm.de
hitradio-ohr.de
neckaralblive.de
dasneueradioseefunk.de

radioton.de
antenne.de
megaradio.bayern
rt1.de
radio-augsburg.de
fanatsy.de
smartradio.de
top-fm.de
christlichesradio.de
mk-online.de
radio2day.de
radioarabella.de
charivari.de
feierwerk.de
radiogong.de
camillo929.de
hitradion1.de
jazzstudio.de
meinlieblingsradio.de
n904beat.de
radiopray.de
aref.de
radiof.de
radio-meilensteine.de
starfm.de
gongfm.de
radio-opera.de
radio8.de
radioawn.de
radio-trausnitz.de
unserradio.de
bayernwelle.de
isw.fm
alpenwelle.de
radio-in.de
radio-oberland.de
extra-radio.de
radio-bamberg.de
euroherz.de
mainwelle.de
radio-plassenburg.de
ramasuri.de
rsa-radio.de
allgaeuhit.de
radiohastagplus.de
radioprimaton.de
primavera24.de
6rtl.com

spreeradio.de
radio-potsdam.de
jam.fm
rs2.de
radio-cottbus.de
bbradio.de
kissfm.de
berliner-rundfunk.de
domradio.de
lausitzwelle.de
fluxfm.de
hitradio-skw.de
jazzradio.net
radio.de
pure-fm.de
radiogold.de
schlager.radio
metropolfm.de
rockland.de
radio21.de
80s80s.de
917xfm.de
hamburg-zwei.de
radiohamburg.de
harmonyfm.de
ffh.de
antennemv.de
ostseewelle.de
ffn.de
meerradio.de
radio38.de
radio90vier.de
radio-hannover.de
radio-nordseewelle.de
radioosnabrueck.de
antenne-ac.de
antennedueselldorf.de
antennemuenster.de
antenneniederrhein.de
antenneunna.de
hellwegradio.de
radio901.de
raidoberg.de
radiobielefeld.de
radiobochum.de
radiobonn.de
radioduisburg.de
radioemscherlippe.de

radioenneperuhr.de
radioerft.de
radioessen.de
radioeuskirchen.de
radioguetersloh.de
radiohagen.de
radioherford.de
radioherne.de
radiohochstift.de
radiokiepenkerl.de
radiokoeln.de
radiokw.de
radioleverkusen.de
radiolippe.de
lippewelle.de
radiomk.de
radiomuelheim.de
radioneandertal.de
radiooberhausen.de
radiorsg.de
radiorst.de
radiorur.de
radiosauerland.de
radiosiegen.de
radiovest.de
radiowaf.de
radiowestfalica.de
radiowmw.de
radiowuppertal.de
welleniederrhein.de
rpr1.de
antenne-io.de
antenne-kh.de
antenne-kl.de
antenne-koblenz.de
antenne-landau.de
antenne-mainz.de
antenne-pirmasens.de
antenne-zweibruecken.de
cityradio-trier.de
classicrock-radio.de
cityradio-saarland.de
salue.de
apolloradio.de
radiochemnitz.de
radiodresden.de
radioerzgebirge.de
radiolausitz.de

radioleipzig.de
radiozwickau.de
vogtlandradio.de
secondradio.de
radiobrocken.de
radiosaw.de
deltaradio.de
rsh.de
antenne-sylt.de
antennethueringen.de
landeswelle.de
radiotop40.de
bermudafunk.org
radio-fds.de
freies-radio.de
freies-radio-wiesental.de
querfunk.de
rdl.de
freefm.de
sthoerfunk.de
wueste-welle.de
lora924.de
radiomuenchen.net
radio-z.net
88vier.de
kcrw.com
medialabnord.de
fsk-hh.org
tidenet.de
antennebergstrasse.de
freies-radio-kassel.de
radiodarmstadt.de
radio-quer.de
radio-rheinwelle.de
radio-r.de
radio-rum.de
radiox.de
rundfunk-meissner.org
lohro.de
nb-radiotreff.de
radio98eins.de
studio-malchin.de
emsvechtewelle.de
leinehertz.net
oeins.de
osradio.de
radio-aktiv.de
radio-jade.de

radio-marabu.de
okerwelle.de
ostfriesland.de
tonkuhle.de
zusa.de
stadtradio-goettingen.de
antenne-bethel.de
muenster.org
coloradio.org
radioblau.de
radiocorax.de
radio-hbw.de
oksh.de
radio-enno.de
radio-frei.de
tu-ilmenau.de
radiolotte.de
radio-okj.de
srb.fm
wartburgradio.org
horads.de
radioaktiv.org
kit.edu
maxneo.de
funklust.de
kanal-c.net
couchfm.de
bonn.fm
campusfm.info
ctdasradio.de
eldoradio.de
hertz879.de
hochschulradio-aachen.de
hochschulradio.de
koelncampus.com
triquency.de
radius921.de
radio-mittweida.de
radiomephisto.de
20min.ch
friday-magazine.ch
24heures.ch
3plus.tv
4plus.tv
5plus.tv
6plus.tv
7radio.ch
smartradio.ch

aargauerzeitung.ch
televox.ch
fcsg.ch
alf-tv.ch
alpenlandtv.ch
alpen-welle.ch
altamega.ch
annabelle.ch
anzeiger-luzern.ch
anzeigerverband-bucheggberg-wasseramt.ch
anzeigergls.ch
anzeigerinterlaken.ch
anzeiger-kirchberg.ch
anzeigerkonolfingen.ch
anzeigermichelsamt.ch
azoe.ch
anzeigerbern.ch
anzeiger-erlach.ch
anzeigertgo.ch
anzeigerverbandbern.ch
anzeigervomrottal.ch
archebdo.ch
arcmusique.ch
arcinfo.ch
auftanken.TV
zulu-media.net
badenertagblatt.ch
bantigerpost.ch
baernerbaer.ch
bazonline.ch
beautyundlife.ch
beobachter.ch
bernerlandbote.ch
berneroberlaender.ch
bestvision.tv
bielbienne.com
bielertagblatt.ch
bilan.ch
handelszeitung.ch
birsfelderanzeiger.ch
bibo.ch
blick.ch
rjb.ch
boostertv.ch
bote.ch
wohleranzeiger.ch
brunni.ch
bzbasel.ch

bernerzeitung.ch
langenthalertagblatt.ch
canal29.ch
canal9.ch
canalalpha.ch
multivideo.ch
absolutenetworks.ch
channel55.tv
chiassotv.ch
chtv.ch
mediaprofil.ch
energy.ch
beviacom.ch
coopzeitung.ch
cdt.ch
CountryRadio.ch
cransmontana.ch
dasmagazin.ch
databaar.ch
derbund.ch
landanzeiger.ch
landbote.ch
azmedien.ch
unter-emmentaler.ch
diaspora-tv.ch
die-neue-zeit-tv.ch
woz.ch
dieutv.com
diisradio.ch
dorfblitz.ch
dorfheftli.ch
dregion.ch
drita.tv
dukascopy.tv
brandsarelive.com
electroradio.fm
engadinerpost.ch
ibexmedia.ch
ewgoms.ch
femina.ch
sonntag.ch
fuw.ch
radiofm1.ch
frauenfelderwoche.ch
freiburger-nachrichten.ch
frequencebanane.ch
fridolin.ch
frutiglaender.ch

startv.ch
gds.fm
effingermedien.ch
generationfm.ch
genevalatina.ch
ghi.ch
volketswilernachrichten.ch
glattwerk.ch
globalfm.ch
glueckspost.ch
grenchnertagblatt.ch
gvfm.ch
hockeyfanradio.ch
hockeyradio-ticino.ch
hoefner.ch
hoengger.ch
hotelradio.fm
caffe.ch
illustrazione.ch
tvtservices.ch
ipmusic.ch
jamesfm.ch
journaldemorges.ch
lejds.ch
jump-tv.ch
jungfrauzeitung.ch
kanal8610.org
kfn-ag.ch
ktfm.ch
lokalinfo.ch
lacote.ch
lagazette.ch
la-gazette.ch
lagruyere.ch
laliberte.ch
laregion.ch
laregione.ch
latele.ch
ladiesdrive.tv
laupenanzeiger.ch
lausannecites.ch
lematin.ch
lemessager.ch
lenouvelliste.ch
lqj.ch
leregional.ch
letemps.ch
lemanbleu.ch

leutv.ch
lfmtv.ch
liewo.li
illustre.ch
limmattalerzeitung.ch
informatore.net
linoradio.com
linthzeitung.ch
btv-kreuzlingen-steckborn.ch
ossingen.tv
loly.ch
lounge-radio.ch
luzernerzeitung.ch
magicradio.ch
marcandcoradio.com
marchanzeiger.ch
maxtv.ch
maxxima.org
tcr-media.com
migrosmagazin.ch
mkd-music.tv
mtv.ch
murtenbieter.ch
musig24.tv
muttenzeranzeiger.ch
mysports.ch
my105.ch
all3media.com
nfz.ch
nzz.ch
nick.ch
niederaemter-anzeiger.ch
nrtv.ch
oberaargauer.ch
oberbaselbieterzeitung.ch
obersee-nachrichten.ch
oberwiggertaler.ch
oltnertagblatt.ch
onefm.ch
onetv.ch
onedancefm.ch
openbroadcast.ch
kabelfernsehen.ch
ne84.ch
rsi.ch
rtr.ch
rts.ch
srf.ch

swissinfo.ch
puls8.ch
powerup.ch
premiumshopping.tv
pressetv.ch
prosieben.ch
radio1.ch
radio20coeurs.net
radio24.ch
radiobeo.ch
radio32.ch
3fach.ch
r3i.ch
argovia.ch
basilisk.ch
radiobern1.ch
radio-bollwerk.ch
radiobus.fm
canal3.ch
radiocentral.ch
radiochablais.ch
radiocite.ch
django.fm
auborddeleau.radio
eviva.ch
radiofr.ch
radiofd.ch
radiogloria.ch
grrif.ch
radiogwen.ch
radioinside.ch
kaiseregg.ch
kanalk.ch
lafabrik.ch
radiolac.ch
lfm.ch
erf.ch
radiololfm.ch
lora.ch
radio-lozärn.ch
radioluzernpop.ch
radiomaria.ch
radiomelody.ch
radiomultikulti.ch
radiomunot.ch
neo1.ch
mitternachtsruf.ch
radioonyx.org

radiopilatus.ch
pinkradio.com
radiopositive.ch
rabe.ch
radioradius.ch
rasa.ch
rhonefm.ch
rro.ch
rouge.com
stadtfilter.ch
suedostschweiz.ch
radiosummernight.ch
sunshine.ch
radioswissclassic.ch
radioswissjazz.ch
radioswisspop.ch
radiotell.ch
radioticino.com
radiotop.ch
verticalradio.ch
radiovolare.com
radiovostok.ch
radiox.ch
radio.ch
radio4tng.ch
radiochico.ch
radio-jazz.com
radiologisch.ch
radioreveil.ch
radiosport.ch
rapbeatsradio.com
redlineradio.ch
ref-gais.ch
zueriost.ch
regiotvplus.ch
restorm.com
rheinwelten.ch
riehener-zeitung.ch
rtvislam.com
rundfunkpositiv.ch
rundfunk.fm
s1tv.ch
sarganserlaender.ch
sat1.ch
bockonline.ch
shf.ch
shn.ch
schweiz5.ch

schweizamwochenende.ch
schweizerfamilie.ch
schweizerhockeyradio.ch
schweizer-illustrierte.ch
skuizz.com
sowo.ch
solothurnerzeitung.ch
sonntagszeitung.ch
radioluz.ch
spoonradio.com
tagblatt.ch
stadtanzeiger-olten.ch
stadt-anzeiger.ch
deinsound.ch
sunradio.ch
surentaler.ch
surprise.ngo
swiss1.tv
upstream-media.ch
swissquote.ch
syri.tv
tagblattzuerich.ch
tagesanzeiger.ch
tele1.ch
tele-d.ch
telem1.ch
tvo-online.ch
toponline.ch
telez.ch
telebaern.ch
telebasel.ch
telebielingue.ch
teleclub.ch
telenapf.ch
tele-saxon.ch
teleswizz.ch
teleticino.ch
televersoix.ch
televista.ch
tep.ch
tessinerzeitung.ch
thuneramtsanzeiger.ch
thunertagblatt.ch
toxic.fm
tdg.ch
mynmz.ch
tvoberwallis.tv
tv-rheintal.ch

telesuedostschweiz.ch
tv24.ch
tv25.ch
tv4tng.ch
tvm3.ch
uristier.ch
sevj.ch
verniervisions.ch
vibracionlatina.com
radiovintage.ch
virginradiohits.ch
meteonews.ch
wiggertaler.ch
willisauerbote.ch
wochenblatt.ch
wochen-zeitung.ch
zofingertagblatt.ch
zugerpresse.ch
zuonline.ch
zsz.ch
10tv.com
6abc.com
abc13.com
abc15.com
abc7.com
abc7chicago.com
abc7ny.com
abcactionnews.com
abcn.ws
aljazeera.com
antena3.com
atresplayer.com
bfmtv.com
boston25news.com
cadenaser.com
canal-plus.com
canal.fr
cbs.com
cbsloc.al
cbslocal.com
cbsn.ws
cbsnews.com
channel7breakingreport.live
cnbc.com
cnn.com
cnn.it
cope.es
cuatro.com

europe1.fr
firstcoastnews.com
fox.com
fox13news.com
fox13now.com
fox2now.com
fox4kc.com
fox59.com
fox6now.com
fox8.com
fox9.com
foxbusiness.com
foxla.com
foxnews.com
globo.com
goodmorningamerica.com
insideedition.com
itv.com
komonews.com
lasexta.com
lbc.co.uk
local10.com
local12.com
mediaset.it
msnbc.com
nbc.com
nbcchicago.com
nbcdfw.com
nbclosangeles.com
nbcnews.com
nbcnews.to
nbcnewyork.com
ndtv.com
news12.com
news4jax.com
news5cleveland.com
ondacero.es
primocanale.it
prosieben.at
rac1.cat
radioclassique.fr
radioitalia.it
rtl.fr
sky.com
sky.it
skytg24news.it
telecinco.es
telemundo.com

tf1.fr
tgcom24.it
weau.com
wpxi.com
wsbtv.com
wxyz.com
24economia.com
affaritaliani.it
agoravox.fr
arcamax.com
bento.de
blitzquotidiano.it
breaknotizie.com
businessinsider.com
businessinsider.de
buzzfeed.com
buzzfeednews.com
caffeinamagazine.it
cnews.fr
ctxt.es
economiadigital.es
elboletin.com
elconfidencial.com
elconfidencialdigital.com
eldiario.es
eldigitalcastillalamancha.es
elespanol.com
estrelladigital.es
fanpage.it
finanzen.net
fivethirtyeight.com
francesoir.fr
gasteizhoy.com
huffingtonpost.co.uk
huffingtonpost.com
huffingtonpost.es
huffingtonpost.fr
huffingtonpost.it
huffpost.com
infolibre.es
kentonline.co.uk
lainformacion.com
lapresse.it
lavozdelsur.es
leggioggi.it
lettoquotidiano.it
libertaddigital.com
linternaute.com

linternaute.fr
livenewsnow.com
livesicilia.it
mediapart.fr
naciodigital.cat
news-mondo.it
news-und-nachrichten.de
news.com.au
news.de
news64.net
newsbreakapp.com
newser.com
newsmondo.it
newsnow.co.uk
nextquotidiano.it
noticias24.com
notizie.it
politico.com
presseportal.de
publico.es
quifinanza.it
quotidianodiragusa.it
racocatala.cat
realclearpolitics.com
republica.com
rosenheim24.de
roughlyexplained.com
salon.com
slate.com
slate.fr
strettoweb.com
tempi.it
termometropolitico.it
theconversation.com
thedailybeast.com
theperspective.com
timesofisrael.com
uol.com.br
valenciaplaza.com
vice.com
vilaweb.cat
vox.com
vozpopuli.com
wired.it
worldjusticenews.com
abc.net.au
ard-text.de
ardmediathek.de

bbc.co.uk
bbc.com
cbc.ca
ccma.cat
channel4.com
france.tv
france24.com
france3.fr
francebleu.fr
franceculture.fr
franceinter.fr
francetelevisions.fr
francetv.fr
francetvinfo.fr
heute.de
npr.org
orf.at
pbs.org
rai.it
rainews.it
raiplay.it
raiplayradio.it
rtbf.be
rte.ie
rtve.es
sbs.com.au
swr.de
tagesschau.de
tv5monde.com
uktv.co.uk
voanews.com
wdr2.de
abc.es
actu.fr
actualites-la-croix.com
adnkronos.com
ansa.it
apnews.com
ara.cat
avvenire.it
baltimoresun.com
bergamopost.it
bloomberg.com
bostonglobe.com
canarias7.es
capital.fr
cataniatoday.it
challenges.fr

chicagotribune.com
corriere.it
corriereadriatico.it
courrierdelouest.fr
courrierinternational.com
dallasnews.com
daytondailynews.com
democratandchronicle.com
denverpost.com
derbytelegraph.co.uk
diaridegirona.cat
diaridetarragona.com
diariocordoba.com
diariodeibiza.es
diariodeleon.es
diariodemallorca.es
diariodenavarra.es
diariodesevilla.es
diarioinformacion.com
diariojaen.es
diariolibre.com
diariosur.es
diariovasco.com
ecodibergamo.it
economist.com
el-nacional.com
elcomercio.es
elcorreo.com
elcorreogallego.es
elcorreoweb.es
eldia.es
eldiariomontanes.es
eleconomista.es
elmundo.es
elnortedecastilla.es
elpais.com
elperiodico.com
elperiodicoextremadura.com
elpuntavui.cat
eltiempo.com
estrepublicain.fr
euro-actu.fr
europapress.es
expansion.com
expressandstar.com
forbes.com
fortune.com
ft.com

gazzettadelsud.it
gazzettadiparma.it
giornaledibrescia.it
giornaledilecco.it
giornaledimonza.it
giornaletrentino.it
giornalone.it
granadahoy.com
heraldo.es
hoy.es
huelvainformacion.es
humanite.fr
ideal.es
ilcorrieredellacitta.com
ilfattoquotidiano.it
ilfoglio.it
ilgazzettino.it
ilgiornale.it
ilgiorno.it
ilmanifesto.it
ilmattino.it
ilmessaggero.it
ilsecoloxix.it
ilsole24ore.com
iltempo.it
independent.co.uk
indiatimes.com
inews.co.uk
internazionale.it
jpost.com
kieler-nachrichten.de
la-croix.com
lagacetadesalamanca.es
lagazzettadelmezzogiorno.it
lamarea.com
lanouvellerepublique.fr
lanuevacronica.com
laopinioncoruna.es
laopiniondemalaga.es
laopiniondemurcia.es
laopiniondezamora.es
lapresse.ca
laprovence.com
laprovincia.es
larazon.es
larepublica.pe
larioja.com
lasprovincias.es

lastampa.it
latimes.com
latribune.fr
lavanguardia.com
laverdad.es
lavoixdunord.fr
lavozdegalicia.es
lavozdigital.es
ledauphine.com
lefigaro.fr
lemonde.fr
lep.co.uk
lepoint.fr
lepopulaire.fr
lesechos.fr
lexpress.fr
liberation.fr
liberoquotidiano.it
lincolnshirelive.co.uk
lne.es
malagahoy.es
marianne.net
mercurynews.com
miamiherald.com
milanotoday.it
motherjones.com
naiz.eus
newsday.com
newsok.com
newsweek.com
nouvelobs.com
nymag.com
nytimes.com
observer-reporter.com
oggitreviso.it
oregonlive.com
orlandosentinel.com
ouest-france.fr
panorama.it
paris-normandie.fr
plymouthherald.co.uk
post-gazette.com
quotidiano.net
quotidianodipuglia.it
quotidianopiemontese.it
regio7.cat
repubblica.it
republicain-lorrain.fr

reuters.com
riminitoday.it
romatoday.it
seattletimes.com
sfgate.com
startribune.com
sudinfo.be
sudouest.fr
sun-sentinel.com
telegraph.co.uk
theatlantic.com
theguardian.com
thehill.com
thetelegraphandargus.co.uk
thetimes.co.uk
time.com
torinotoday.it
trevisotoday.it
udinetoday.it
usnews.com
valeursactuelles.com
washingtonexaminer.com
washingtonpost.com
washingtontimes.com
waz.de
wired.com
wsj.com
20minutes.fr
20minutos.es
20mn.fr
blick.de
dailymail.co.uk
dailyrecord.co.uk
dailystar.co.uk
eveningtimes.co.uk
express.co.uk
hulldailymail.co.uk
ilrestodelcarlino.it
journaldemontreal.com
lanazione.it
leggo.it
leparisien.fr
metro.co.uk
metro.it
mirror.co.uk
nydailynews.com
nypost.com
parismatch.com

que.es
standard.co.uk
suntimes.com
thesun.co.uk
trinitymirror-news.co.uk
usatoday.com
watson.ch
watson.de
weltwoche.ch
3sat.de
arte.tv
bb-mv-lokaltv.de
br.de
daserste.de
dw.com
hr-fernsehen.de
kika.de
mdr.de
ndr.de
phoenix.de
radiobremen.de
rbb-online.de
sr.de
swrfernsehen.de
ard.de
zdf.de
deutschlandfunk.de
deutschlandfunkkultur.de
deutschlandradio.de
deutschlandfunknova.de
hr1.de
hr2.de
hr3.de
hr4.de
you-fm.de
hr-inforadio.de
n-joy.de
antennebrandenburg.de
fritz.de
inforadio.de
radioeins.de
unserding.de
dasding.de
swr3.de
wdrmaus.de
rsi.ch
rtr.ch
rts.ch

srf.ch
swissinfo.ch

**Appendix B: Automated content analysis approach for political information detection**

To design a dictionary for detecting political information, we extracted terms from the German codebook for the Comparative Agendas Project (CAP; Bevan 2019). The codebook is used to label political themes in Germany (e.g. economy or foreign politics) and includes key terms related to them. We also added a theoretically conceptualized list of terms for topics underrepresented in the CAP codebook (i.e. elections and ecology), together with a list of political actors' names in Germany/Switzerland (e.g. members of parliament) and G20/EU countries (e.g. presidents and vice-ministers).

To prepare the two validation sets for political content classification, coders manually labeled two sets of documents. The first set of documents consisted of 594 documents, namely short (Twitter and Telegram posts by politicians and journalists) and longer documents (articles from Germanophone legacy media, such as Süddeutsche Zeitung, and right-wing outlets, such as Journalistenwatch) crawled online. The second set consisted of 262 documents randomly sampled from the tracking data. For a summary of this process, see Table B1.1 below.

**Table B1.1**
*Summary of the procedures for preparing the validation datasets*

| Dataset name | Size | Source | Preparation | Coding validation |
| --- | --- | --- | --- | --- |
| Diverse validation dataset | 594 documents | Short (Twitter and Telegram posts by politicians and journalists) and longer documents (articles from German legacy media, such as *Süddeutsche Zeitung*, and right-wing outlets, like *Journalistenwatch*) crawled online | 4 coders | 10% of the dataset used for reliability check: Cohen's Kappa = .86 (political actor) and Cohen's Kappa = .68 (political topic); complete dataset checked and consensus-coded by 2 experts |
| Web-tracking validation dataset | 262 documents | Random sample of web-tracking data | 1 coder | For all cases where the manual coding was inconsistent with the classification, the coding was reviewed, discussed, and consensus-coded by 3 experts |

To determine whether the document is related to politics, we calculated the ratio between the number of politics-related unique terms to the overall number of unique terms per document. After calculating all the ratios per validation dataset, we then used each of these ratios as a possible threshold for assigning a political label to all the documents in the respective validation dataset and calculated the resulting average F1 scores. Then, we chose the threshold that achieved the maximum F1 score for the web-tracking validation dataset due to it being the closest in terms of quality and composition to our data. The best performance under these circumstances was achieved with the threshold of at least 0.25% of all unique terms in the document being present in the political dictionary.

Following the comparison of different modes of pre-processing (e.g. stemming, lemmatisation, stop word removal), we opted for stemming (using snowball stemmer from NLTK Python library) together with the selectolax HTML parser to extract text from tracked HTML content. The decision was based on the computational time required to process the entire tracking sample per mode of pre-processing and the impact of specific modes of pre-processing on dictionary performance. Because of the classifier's low performance on extremely short texts, documents with less than 1,000 characters were filtered out of the tracking dataset before the classifier was applied. The manual examination of these short documents showed that they did not contain meaningful information and usually consisted of error messages (e.g. about the page not loading).

Each document in the two sets was coded according to the following three variables: 1) **POLITICAL ACTOR, POLITICAL INSTITUTIONS AND PRINCIPLES (i.e. *Are political actors mentioned in the text? Does the text mention political institutions?)*** ; 2) **POLITICAL TOPICS (*Are political topics/issues mentioned in the text?)***; 3) **OTHER POLITICAL SUBJECT AREA.** The detailed description of the variables is provided below; the document was classified as political if at least one of these variables was present in the document.

**Variable: POLITICAL ACTOR, POLITICAL INSTITUTIONS AND PRINCIPLES**

***Are political actors mentioned in the text? Does the text mention political institutions?***

0 = no

1 = yes

*municipal / regional / national / international political actors, i.e.*

- (members of) governments, e.g. government, president, chancellor, minister
- (members of) governments, e.g. government, president, chancellor, minister
- opposition, e.g. opposition leader, opposition parliamentary group
- Parliamentarian, e.g. (member of) the Bundestag, (member of the Parliament), Nationalrat, National Councilor
- politician
- parties, e.g. party leader, party members
- administrations, e.g. ministries

  *politically active citizens who take a clear political action, e.g. as a*

- lobbying organization
- formulation demands on politicians
- organize demonstrations

  *entire states as political actors*

  *local / regional / national / international political institutions and principles, e.g.*

- constitution,
- political institutions, e.g. parliament, Council of Europe, European Union, courts of law
- laws and agreements
- political culture (e.g. political attitudes of citizens)

**Variable: POLITICAL TOPICS**

*Are political topics/issues mentioned in the text?*

0 = no

1 = yes

*political topics/issues at the municipal / regional / national / international level, e.g.:*

- working conditions and labor market
- foreign and domestic trade
- banks and finances
- education
- civil rights and liberties, minority rights
- German reunification
- energy
- family and social affairs
- health care
- international relations and foreign aid
- agriculture
- mobility, transport, and traffic
- public administration
- political campaigns, referendums, initiatives, votes
- law and crime
- environment
- defense
- water management
- housing, construction, and spatial planning
- economy

- science and technology

- other subject area ▸ if applicable: name in variable 'OTHER POLITICAL SUBJECT AREAS'

A. Content from these subject areas is political if it presents problems / proposed solutions from a perspective relevant to a population group or society as a whole (not just to an individual).

B. Political issues can be described at different stages of the process:

- As a pure description of the problem (without a policy demand or action).

- As a demand for policy (still without action).

- As political discussions or actions.

**Variable: OTHER POLITICAL SUBJECT AREAS**

*Information deals with a political content in a subject area that was not mentioned in the variable 'POLITICAL TOPICS'*

- Name of the subject area.